\documentclass[12pt]{article}
\usepackage{graphics,epsfig}
\usepackage{graphicx}
\usepackage{float}
\usepackage{amssymb,amstext,amsmath}
\usepackage{mathtools}
\usepackage{booktabs}
\usepackage[latin1]{inputenc}
\usepackage{epstopdf}

\begin{document}
\title{Two universal laws for plastic flows and the consistent thermodynamic dislocation theory} 
\author{K.C. Le\,$^{a,b}$\footnote{E-mail: lekhanhchau@tdtu.edu.vn}}
\date{$^a$\,Materials Mechanics Research Group, Ton Duc Thang University, Ho Chi Minh City, Vietnam\\
$^b$\,Faculty of Civil Engineering, Ton Duc Thang University, Ho Chi Minh City, Vietnam}
\maketitle
\centerline{\bf Abstract}
This paper verifies two laws for plastic flows of face-centered cubic crystals that deform at constant strain rates and fixed ambient temperatures. The first law relates steady-state flow stress to ambient temperature and strain rate. The second law requires an increase of configurational entropy towards a maximum reached in the steady state. The large scale least squares analysis is provided which allows the physics-based parameters of thermodynamic dislocation theory to be identified in accordance with these laws.  


\bigskip

It is experimentally observed that a dislocated crystal deforming at constant strain rate and fixed ambient temperature will approach a steady state of plastic flow, and the corresponding steady-state flow stress, $\sigma_s$, depends on the ambient temperature $T$ and the strain rate $\dot{\varepsilon}$. Kocks and Mecking \cite{KM-03} were the first to formulate the following universal law for the plastic flow of fcc-crystals: The steady-state flow stress is a function of the combination of ambient temperature and strain rate, $(T/T_P) \ln (\dot{\varepsilon}_r/\dot{\varepsilon})$. Here $T_P$ is an energy barrier expressed in the temperature unit, while $\dot{\varepsilon}_r$ is a reference strain rate. However, the empirical quadratic function proposed in \cite{KM-03}, which contains the square root of this combination, is not appropriate for two reasons: (i) this function does not fit the experimentally observed steady-state flow stresses, which are usually greater than those obtained by extrapolation based on the Voce law, (ii) it cannot be derived from the first principle calculation. The alternative scaling law for the steady-state flow stress can be obtained from the kinetics of thermally activated dislocation depinning first proposed by Langer, Bouchbinder and Lookman \cite{LBL-10}. Applying the inverse relationship to the double exponential formula for the plastic strain rate (see Eq.~(5.4) in \cite{LBL-10}) to the steady state, the following scaling law is obtained 
\begin{equation}
\label{eq:1}
\frac{\sigma_s}{\sigma_{Ts}}=\ln \Bigl( \frac{1}{\frac{T}{T_P}\ln (\frac{\dot{\varepsilon}_r}{\dot{\varepsilon}})}\Bigr).
\end{equation}
Here, $\sigma_{Ts}=\mu (T) \alpha b\sqrt{\rho_s}$ is the steady-state Taylor stress, $\mu(T)$ the shear modulus that depends on the ambient temperature, $b$ the Burgers' vector, $\rho_s$ the steady-state dislocation density, and $\dot{\varepsilon}_r=b\sqrt{\rho_s}/t_0$, where $t_0$ is the time characterizing the depinning rate. It must be emphasized that \eqref{eq:1} is derived under the assumption that the depinning rate, by being the slow ``bottleneck'', is dominant, and that therefore the time for dislocations to move between pinning sites and specific effects such as cross slip could be neglected. The other main assumption is that the energy barrier and the steady-state dislocation density are independent of strain rate and temperature. The scaling law \eqref{eq:1} provides the method for determining  the three material parameters $s=\alpha b\sqrt{\rho_s}$, $T_P$, and $\dot{\varepsilon}_r$ from the experimental data. To the author's knowledge, this has not yet happened, so it remains unclear whether this law is supported by the experiment and in what temperature and strain rate range it is valid. To clarify this matter I use the accurate data obtained from the compression tests of copper (aluminum) at three (four) different elevated temperatures and four (five) different strain rates \cite{SSK-71}, with the quasi-static case being excluded, and identify that, for pure copper, $s=6.3915 \times 10^{-3}$, $T_P=45000\,$K, $\dot{\varepsilon}_r=3.16\times 10^{12}/$s, while for pure aluminum, $s=5.4526 \times 10^{-3}$, $T_P=27800\,$K, $\dot{\varepsilon}_r=7.5\times 10^{11}/$s. As discussed in \cite{LBL-10}, the kinetics of dislocation depinning breaks down at the very low temperature and small strain rate limits. This is the reason why I excluded the quasi-static case in this scaling analysis. Note also that the shear modulus depends on the ambient temperature according to $\mu(T)=\mu_1 - D/(\exp(T_1/T)-1)$, where $\mu_1 = 51.3\,$GPa, $D = 3\,$GPa, $T_1 = 165\,$K for copper, and $\mu_1 = 28.8\,$GPa, $D = 3.44\,$GPa, $T_1 = 215\,$K for aluminum (see \cite{VARSHNI-70}).

\begin{figure}[htb]
\centering \includegraphics[height=7cm]{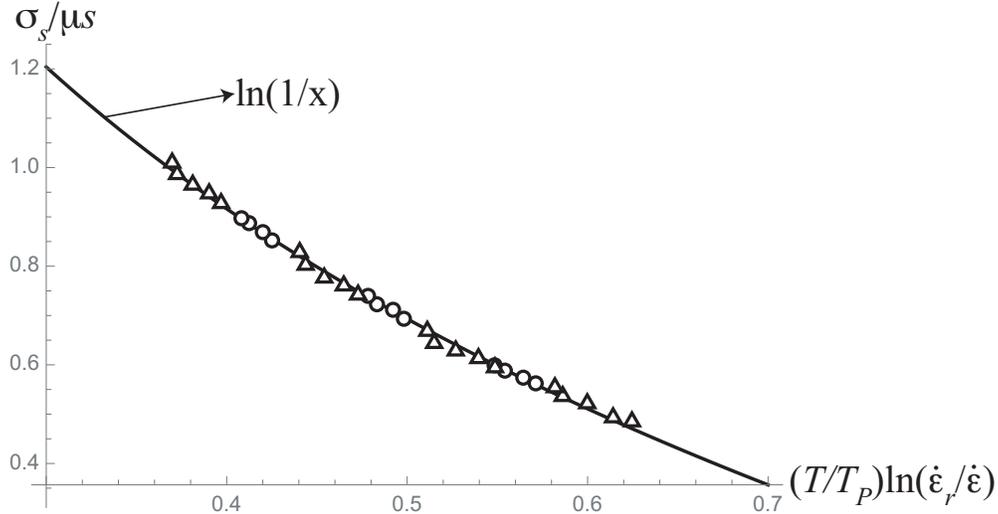} \caption{Dimensionless steady-state flow stresses $\frac{\sigma_s}{\mu s}$ versus $\frac{T}{T_P}\ln (\frac{\dot{\varepsilon}_r}{\dot{\varepsilon}})$ for copper (circle) and aluminum (triangles) and the master curve $y=\ln(1/x)$.} \label{fig:1}
\end{figure}

Fig.~\ref{fig:1} shows the data points with $x$-coordinate being $(T/T_P) \ln (\dot{\varepsilon}_r/\dot{\varepsilon})$ and $y$-coordinate being $\sigma_s/\mu(T)s$ of copper (circles) and aluminum (triangles) as well as the master curve $y=\ln (1/x)$. It is seen that most points lie almost exactly on this curve. Since the experimental points of other fcc crystals such as silver or nickel are also close to those of copper and aluminum when appropriately scaled \cite{KM-03}, it is concluded that Eq.~\eqref{eq:1} is the validated scaling law for the steady-state flow stress of these materials for temperatures from room temperature to two-thirds of the melting temperature and for strain rates from $1/$s to at least $10^6/$s.

The law \eqref{eq:1}, however, does not say anything about how the stress and dislocation density approach the steady state. This behavior can be extracted from a second universal law for plastic flows formulated also by Langer et al. \cite{LBL-10} as follows: The configurational entropy of the subsystem of dislocations must increase and reach its maximum in the steady state. This law is the consequence of the thermodynamics and statistical mechanics of configurational subsystem of moving dislocations regarded as a dissipative driven system. The underlying thermodynamics is based on the existence of slow and fast variables in this system. Fast variables are coordinates of dislocations. Slow variables are elastic deformation, dislocation density, and configurational entropy (or effective disorder temperature). The conditions under which fast variables can be averaged out are not the same as those of reversible Hamiltonian systems for which ergodicity is crucial \cite{RU-04}. The laws governing the slow variables are also not the same as those of equilibrium thermodynamics of ergodic Hamiltonian systems. Even the steady state, regarded as ``equilibrium'' state of the configurational subsystem, is not a strict equilibrium, since dislocations are permanently pinned and depinned and move between the pinning sites so that the body flows plastically at the constant strain rate. This is similar to the slow change of amplitude of non-linear vibration of a forced dissipative oscillator towards the steady-state amplitude after the fast oscillation is averaged out \cite{GUCK-13,LE-14}. Although the dissipative configurational subsystem of dislocations is driven, it seems physically reasonable that the configurational entropy must increase and reach a maximum in the steady state regarded as ``equilibrium''.

The theory based on the law of maximum configurational entropy was proposed in \cite{LBL-10} and slightly modified in \cite{LE-20} for polycrystals. Its governing equations for the stress $\sigma$, the dislocation density $\rho$, and the  configurational temperature $\chi$, read
\begin{equation}
\label{eq:2}
\begin{split}
\frac{\mathrm{d} \sigma}{\mathrm{d} \varepsilon}=2\mu (1+\nu)\Bigl( 1-\frac{q}{q_0}\Bigr),
\\
\frac{\mathrm{d} \rho}{\mathrm{d} \varepsilon}=\frac{\kappa_\rho }{\mu \alpha \zeta ^2(\rho,q_0,T)b^2} \frac{\sigma q}{q_0}\Bigl[1-\frac{\rho}{\rho_s(\chi)}\Bigr] , 
\\
\frac{\mathrm{d} \chi}{\mathrm{d} \varepsilon}=\frac{\kappa_\chi }{\mu \alpha }\frac{\sigma q}{q_0} \Bigl(1-\frac{\chi}{\chi_0}\Bigr) . 
\end{split}
\end{equation}
The first equation is nothing else but Hooke's law in rate form, where $q=\dot{\varepsilon}^pt_0=(b/t_0)\sqrt{\rho}\exp(-(T_P/T)\exp(-\sigma/\sigma_{T}))$ is the normalized plastic strain rate, while $q_0=\dot{\varepsilon}t_0$ is the normalized total strain rate and $\nu$ the Poisson ratio. The other two equations describe the approach of dislocation density and the configurational temperature to the steady-state values based on the above mentioned second law. Here $\rho_s(\chi)=(1/a^2)\exp(-e_d/\chi)$ is the most probable (steady-state) dislocation density at fixed configurational temperature, while
\begin{equation*}
\zeta(\rho,q_0,T)=\ln \Bigl( \frac{T_P}{T}\Bigr)-\ln\Bigl[\ln\Bigl(\frac{b\sqrt{\rho}}{q_0}\Bigr)\Bigr] .
\end{equation*}

For the purpose of numerical integration and parameter identification it is convenient to introduce the rescaled variables and rewrite the system \eqref{eq:2} in the dimensionless form. To this end, the dimensionless dislocation density and effective disorder temperature are introduced as
\begin{equation*}
\tilde{\rho }=a^2 \rho, \quad \tilde{\chi }=\frac{\chi }{e_d}.
\end{equation*}
With these rescaled quantities the dimensionless steady-state dislocation density at fixed configurational temperature becomes
\begin{equation*}
\tilde{\rho}_s(\tilde{\chi})=e^{-1/\tilde{\chi }}.
\end{equation*}
Let the dimensionless ordinary temperature be $\tilde{\theta} =T/T_P$. Then the normalized plastic strain rate can be written as
\begin{equation*}
q(\sigma,\rho,T)=\dot{\varepsilon}^pt_0=(b/a)\tilde{q}(\sigma,\tilde{\rho},\tilde{\theta}),
\end{equation*}
where
\begin{equation*}
\tilde{q}(\sigma,\tilde{\rho},\tilde{\theta})=\sqrt{\tilde{\rho}}\exp \Bigl[-\frac{1}{\tilde{\theta}} e^{-\sigma/\mu r \sqrt{\tilde{\rho}}} \Bigr] ,\quad r=\alpha (b/a)=s/\sqrt{e^{-1/\tilde{\chi }_0}}.
\end{equation*}
The formula for $\zeta$ becomes
\begin{equation*}
\tilde{\zeta}(\tilde{\rho},\tilde{q}_0,\tilde{\theta})=\ln \Bigl( \frac{1}{\tilde{\theta}}\Bigr)-\ln\Bigl[\ln\Bigl(\frac{\sqrt{\tilde{\rho}}}{\tilde{q}_0}\Bigr)\Bigr] . 
\end{equation*}
Using $\tilde{q}$ instead of $q$ as the dimensionless measure of plastic strain rate, we are effectively rescaling $t_0$ by a factor $a/b$: $\tilde{t}_0=(a/b)t_0$. From the definition of $\dot{\varepsilon}_r$ in Eq.~\eqref{eq:1} we obtain $\tilde{t}_0=\sqrt{\exp(-1/\tilde{\chi}_0)}/\dot{\varepsilon}_r$; and we use $\tilde{q}_0=\tilde{t}_0\dot{\varepsilon}$ for converting from $\tilde{q}_0$ to the measured total strain rates.

In terms of the introduced rescaled variables the governing equations read
\begin{align}
\frac{\mathrm{d} \sigma}{\mathrm{d} \varepsilon}&=2\mu (1+\nu) \Bigl( 1-\frac{\tilde{q}}{\tilde{q}_0}\Bigr), \notag
\\
\frac{\mathrm{d} \tilde{\rho}}{\mathrm{d} \varepsilon}&=K_\rho \frac{\sigma }{\mu r \tilde{\zeta}^2(\tilde{\rho},\tilde{q}_0,\tilde{\theta})}\frac{\tilde{q}}{\tilde{q}_0}\Bigl[1-\frac{\tilde{\rho}}{\tilde{\rho}_s(\tilde{\chi})}\Bigr] , \label{eq:3}
\\
\frac{\mathrm{d} \tilde{\chi}}{\mathrm{d} \varepsilon}&=K_\chi \frac{\sigma }{\mu r}\frac{\tilde{q}}{\tilde{q}_0} \Bigl(1-\frac{\tilde{\chi}}{\tilde{\chi}_0}\Bigr) ,\notag
\end{align}
where $K_\rho =\frac{\kappa_\rho a}{b}$ and $K_\chi = \frac{\kappa_\chi b}{a e_d}$. Based on the same observation as in \cite{LBL-10} I assume that $K_\rho$ is independent of the strain rate and temperature, while $K_\chi=c_0\exp(T/c_1)$, where $K_\rho $, $c_0$ and $c_1$ are material constants.

\begin{figure}[htb]
\centering \includegraphics[width=9cm]{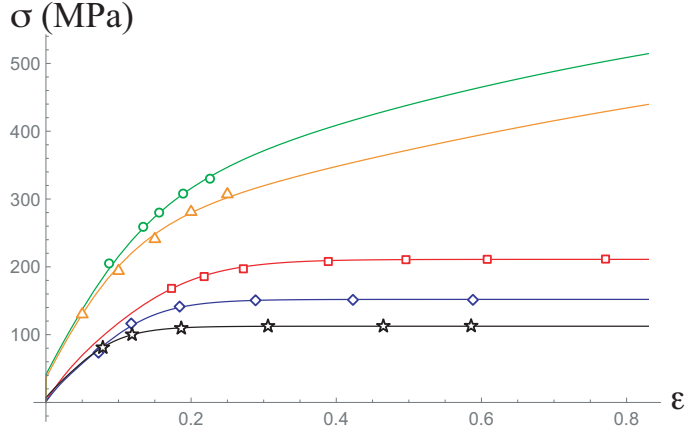} \caption{Stress-strain curves of compression tests for polycrystalline copper. From top to bottom: (i) $T=298\,$K, $\dot{\varepsilon}=9500/$s (green), (i) $T=298\,$K, $\dot{\varepsilon}=81/$s (orange), (iii) $T=873\,$K, $\dot{\varepsilon}=2300/$s (red), (iv) $T=1023\,$K, $\dot{\varepsilon}=960/$s (blue), (v) $T=1173\,$K, $\dot{\varepsilon}=960/$s (black). The data points are taken from \cite{SSK-71,FK-88}. The theoretical curves are simulated in accordance with \eqref{eq:2}. } \label{fig:2}
\end{figure}

\begin{table}[ht]
  \centering 
\begin{tabular}{|c|c|c|c|c|c|}
\hline
$T$ (K), $\dot{\varepsilon}$ (1/s) & $298,9500$ & $298,81$ & $873,2300$ & $1023,960$ & $1173,960$ \\ \hline
$\tilde{\rho}_i$ & $5.43\times 10^{-5}$ & $5.31\times 10^{-5}$ & $1.2\times 10^{-6}$ & $3.08\times 
10^{-7}$ &  $3.24\times 10^{-5}$\\  
\hline
$\tilde{\chi}_i$ & 0.184 & 0.178 & 0.146 & 0.154 & 0.184 \\ \hline
\end{tabular}  
\caption{The initial values of $\tilde{\rho}_i$ and $\tilde{\chi}_i$ for copper}
\label{table:1}
\end{table}

The problem with using system of equations \eqref{eq:3} to simulate stress-strain curves is the choice of parameters. Unfortunately, the choice made in \cite{LBL-10} is not fully consistent with the scaling law \eqref{eq:1}. For instance, the selected value of $T_P=40800\,$K for copper is somewhat smaller than the value $45000\,$K identified from Eq.~\eqref{eq:1}. Similarly, for the ad-hoc selected parameters $\chi_0$, $a$, and $t_0$, it is found that $\dot{\varepsilon}_r=1.35\times 10^{11}/$s which is less than the value $3.16\times 10^{12}/$s identified above. Besides, in contrast to \cite{LBL-10}, the parameters $r$ and $\tilde{t}_0$ need not be assumed or identified at all, because as seen above, they can be computed from $\tilde{\chi}_0$, $s$, and $\dot{\varepsilon}_r$. Therefore the inconsistent and ad-hoc choices made in \cite{LBL-10} are abandoned and all parameters and initial conditions are identified with the large-scale least-squares analysis \cite{LT-17,LTL-17}. This yields in addition to $T_P$, $s$, and $\dot{\varepsilon}_r$ the four basic parameters for copper $\tilde{\chi}_0=0.233$, $K_\rho=1.66$, $c_0=0.4$, $c_1=166\,$K. With this I find that $r=0.055$, $\tilde{t}_0=3.7\times 10^{-14}$s which is also consistent with Eq.~\eqref{eq:1}. Note that, within this theory, only the combination $\tilde{\rho}_s=a^2\rho_s$ can be identified which is equal to $0.01$. If we assume that $a=3b$, then $\rho_s=2.32\times 10^{16}/$m$^2$, and $t_0=1.23\times 10^{-14}$s. Fig.~\ref{fig:2} presents five representative stress-strain curves for copper under compression at five different thermal and loading conditions as results of the simulation of Eq.~\eqref{eq:2} together with the experimental points taken from \cite{SSK-71,FK-88}. Note that the behavior of these curves near the onset of plastic yielding is sensitive to initial dislocation densities and configurational temperatures whose rescaled quantities are presented in Table~\ref{table:1}. Note also that the two upper (green and orange) curves in Fig. ~\ref{fig:2}, taken from \cite{FK-88}, were measured under conditions beyond those of compression tests considered in \cite{SSK-71}, which were used to verify the scaling law \eqref{eq:1}. Nevertheless, the parameters identified with Eq.~\eqref{eq:1} still give good agreement here, even for small strains where the finite yield stresses are roughly of the same magnitude as that in Figure~7 of \cite{FK-88}. These consistencies are significant. The excellent agreement between theory and experiment and the consistencies with the formulated universal laws allow the conclusion that this theory can be used to predict the plastic flows of fcc-crystals over a wide range of strain rates and temperatures.

\bigskip
\noindent {\it Acknowledgement.}

I would like to thank J. S. Langer for the helpful discussion and many valuable suggestions, which considerably improve the first draft of the paper.

\end{document}